# Towards planar Iodine 2D crystal materials

Xinyue Zhang, Qingsong Huang*

## Abstract

Usually, the octet-rule has limited the 2D materials by elements within groups IIIA –VA, whose outmost electrons can be considered to form hybridization orbit by *s* wave and *p* wave. The hybridization orbits can accommodate all the outmost electrons and form robust σ bonds. Over VA group, the outmost electrons seem too abundant to be utilized in hybridization orbits. Here we show a *spd$^2$* hybridization rule, accommodating all the outmost electron of halogen elements. Each atom can be connected with contiguous atom by robust σ bond, and carries one dangling lone-unpaired electron, implying formation of π bond is possible. One atomic iodine layer can be robustly locked by the σ bond, forming iodiene sheet by *spd$^2$* hybridization orbits.

With application of compression strain, the π bond forms, and the band inversion occurs simultaneously at the valence band and conductance band. Dirac points (line) and topological nontrivial points (arc or hoop) suggests the transformation of Dirac semimetal or topological semimetal happens.

## Introduction

Upon 2D materials[1], e.g. graphene, the unpaired dangling electrons to adjacent atoms can form weak π bond, inducing linear-energy dispersion for survival of mass-less Dirac-Fermions[2,3].In particular, main VA group elements, such as phosphorous[4], arsenic[5], antimony[6], and bismuth[7], form elemental 2D materials supported mainly by σ bond, hybridizing by *p* orbitals, with a little bit of *s* orbitals hybridization [8]. The topological nontrivial points or Dirac points arise from appearance of band inversion[9-12], instead of just formation of π bonds[8].The π bond just provide a possibility to linear-energy dispersion of the survival of Dirac-points, supporting the existence of massless Dirac-Fermions[13,14]. By far, the group IIIA-VA are generally regarded as limitation scope for forming

* School of chemical engineering, Sichuan University, Chengdu, 610065, P. R. China
  E-mail: qshuang@scu.edu.cn
  (Prof. Qingsong Huang)

elemental 2D materials by *s-wave* and *p-wave* - hybridization orbits[15].

As for the VIA and VIIA elements, however, formation of elemental 2D materials remains challenging. Since the outmost electrons are more than six, formation of 2D materials need at least 3 covalent bonds, therefore, violation of octet rule is inevitable. The empty *d* orbits should be considered to be hybridized with *s* and *p* orbits to form *spd* hybridization orbits, accommodating the redundancy electrons. The covalent bonds between elements should be strong enough to sustain the frame of the 2D configurations.

Here we present 2D materials arising from element of group VIIA, constructing σ bonds by *spd²* hybridization orbit. Although the *spd* hybridization can survive in some specific molecules[16,17], it has never been observed in infinite 2D periodic crystals before.

So far, some literatures have reported the 3D iodine molecule ($I_2$) crystalline stacked by 2D molecule layers, transferring into covalent-connection atomic layer in-plane and Van der Waals connection inter-plane by pressure-inducing process[18-22]. The $I_2$ molecule crystalline is stacked by atomically thin layers composing of coplanar zigzag networks, where the van der Waals force work in both in-plane and inter plane. While the $I_2$ molecule dissociate into atoms, inducing by high pressure, triggering the in-plane atoms become connection by covalent bonding, where the atoms inter-plane interact by van der Waals force. The pressure-induced covalent bond survives because of *spd* hybridization orbits.

According to the experiment references[23], the iodiene do exist in high pressure. Herewith, we try to construct iodiene in atmosphere pressure, obtaining tetragonal 2D configuration by DFT geometrical optimization. As for tetragonal 2D configuration crystal material, the atoms are connected by *spd²* hybridized σ bonds, while the π key in high symmetry points is difficult to form since the long bond length (LBL), despite the unpaired lone electron is available for each atom[24-26]. Formation of Dirac point depends on band inversion[27], bond length reductions to form π bond[28], or perturbation of Anderson potential barrier[29]. The phonon spectrum calculation shows that iodine can form stable two-dimensional materials (Fig. 1). Provided absence of strain, the LBL is too far to form π key. When strain is applied, the LBL becomes smaller than before, and a Dirac point comes into being because of band inversion.

# Discussion

## 1. Structure and stability

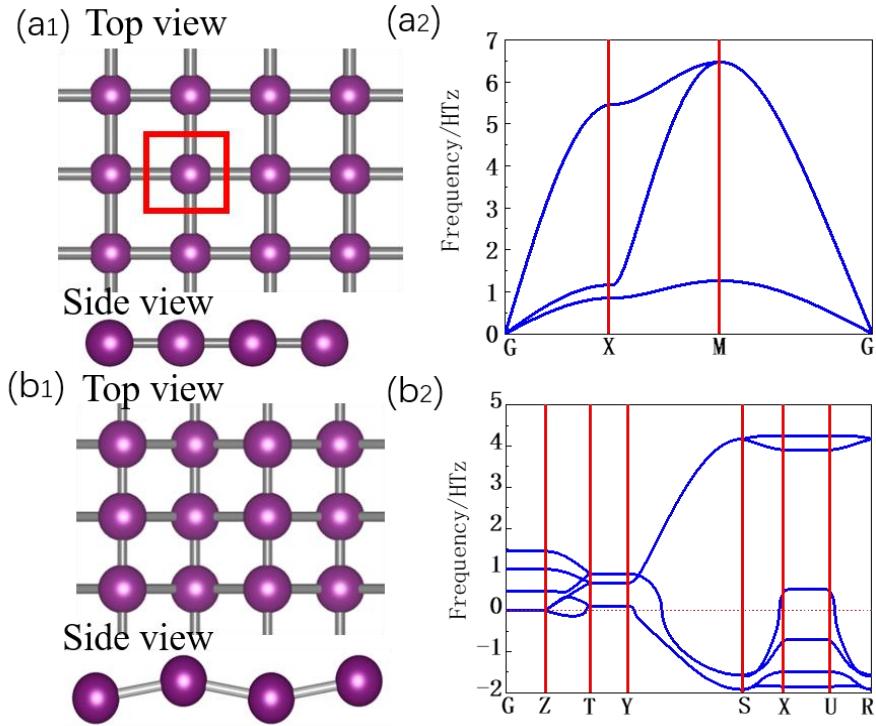

**Fig. 1** Possible structure and phonon spectrum of iodiene. (a$_1$): Structural model illustrates a single layer with a planar configuration and a geometric unit, which is derived from a 3D iodine structure under high pressure. The red frame shows the unit cell of iodiene. (a$_2$): Phonon spectrum shows no virtual frequency is available, implying the planar 2D configuration is stable. (b$_1$): Buckling structural model, including top view and side view. (b$_2$): Phonon spectrum of buckling iodiene demonstrates the existence of virtual frequency, indicating the buckling configuration is unstable.

As we all know, the iodine molecular crystal can be compressed under 20.6 MPa into atomic crystal, stacking up by 2D atomic layer-iodiene, where each atom is connected by covalent bonds *in-plane*. Iodine crystal can gradually transfer from orthorhombic to body-central-cubic[30-32], activating evolution of crystal configuration by high pressure. Since the layered 3D materials are composed of 2D iodiene, coupling between contiguous layers by Van der Waals force. Exfoliating any layer from the body, the isolating layer can become free standing. To this end, a DFT model is built up to simulate the optimized structure, demonstrating a tetragonal iodiene configuration can become really existing (Fig. 1a$_1$ top view). The unit cell has been labeled as red frame, demonstrating bond length is around 3.234 Å, and bond angle is 90 º (Fig. 1a$_1$ side view). No imaginary frequency in phonon spectrum illustrates planar iodiene is stability, even if the ultra-high compression pressure has been relaxed (Fig. 1a$_2$). Details such as total energy and binding energy of iodiene are shown in table S1.

In addition, a buckling spatial configuration can be considered (Fig. 1b$_1$) to be instable, deducing from the negative phonon spectrum of acoustic branch (Fig. 1b$_2$). Moreover, the puckered spatial configuration seems impossible (Fig. S1), because puckered iodiene always develops into a quasi-plane structure after a geometrically optimization. Moreover, the planar tetragonal structure is robust and stable, if specific bond length is set within Van der Waals diameter[33] (Fig. S2).

## 2. Effect of strain on band structure

The band gap can be tuned by biaxial strain[34], including tensile strain and compression strain. When the tensile performed in real space, the band gap can be opened even wider than before (Fig. 2a-b). Just like exerting a pulling separating-force between the conductance bands (CB, Red part) and valence bands (VB, blue part) in BZ. Once applying biaxial compression strain to 2D iodiene in real space, it looks like exerting a pushing closing force between CB and VB in BZ. Thus, the compression strain can induce reduction of band gap (Fig. 2c). Once the compression strain reaches 11.5%, the direct band gap at X point in the Brillouin zone (BZ) approaches zero (Fig. 2d), implying the formation of π bond is possible. Further compression to strain of 12.5% can induce band inversion, making Dirac points or topological non-trivial points available (Fig. 2e). The 2D iodiene can transfer from wide band gap to Dirac semimetal[35,36] or topological semimetal. When the strain increases to 15%, the band inversion becomes even strong near the high symmetry X point (Fig. 2f). The Dirac points or topological nontrivial points can be linked into arcs (Fig. 2g), implying the Dirac semimetal[35,36] or topological semimetal. The inverted band structures can refer to the 3D structures in detail, as shown in Fig. 2h and Fig. 3. The 3D band structures with biaxial stress can be found in Fig. 2h, where the compression strain at 15% has been demonstrated. While the 3D band structure in Fig. 2h has illustrated the contact Dirac points, where the Dirac semimetal and topological semimetal can be viewed. All the high symmetry points can be found in Fig. 2i.

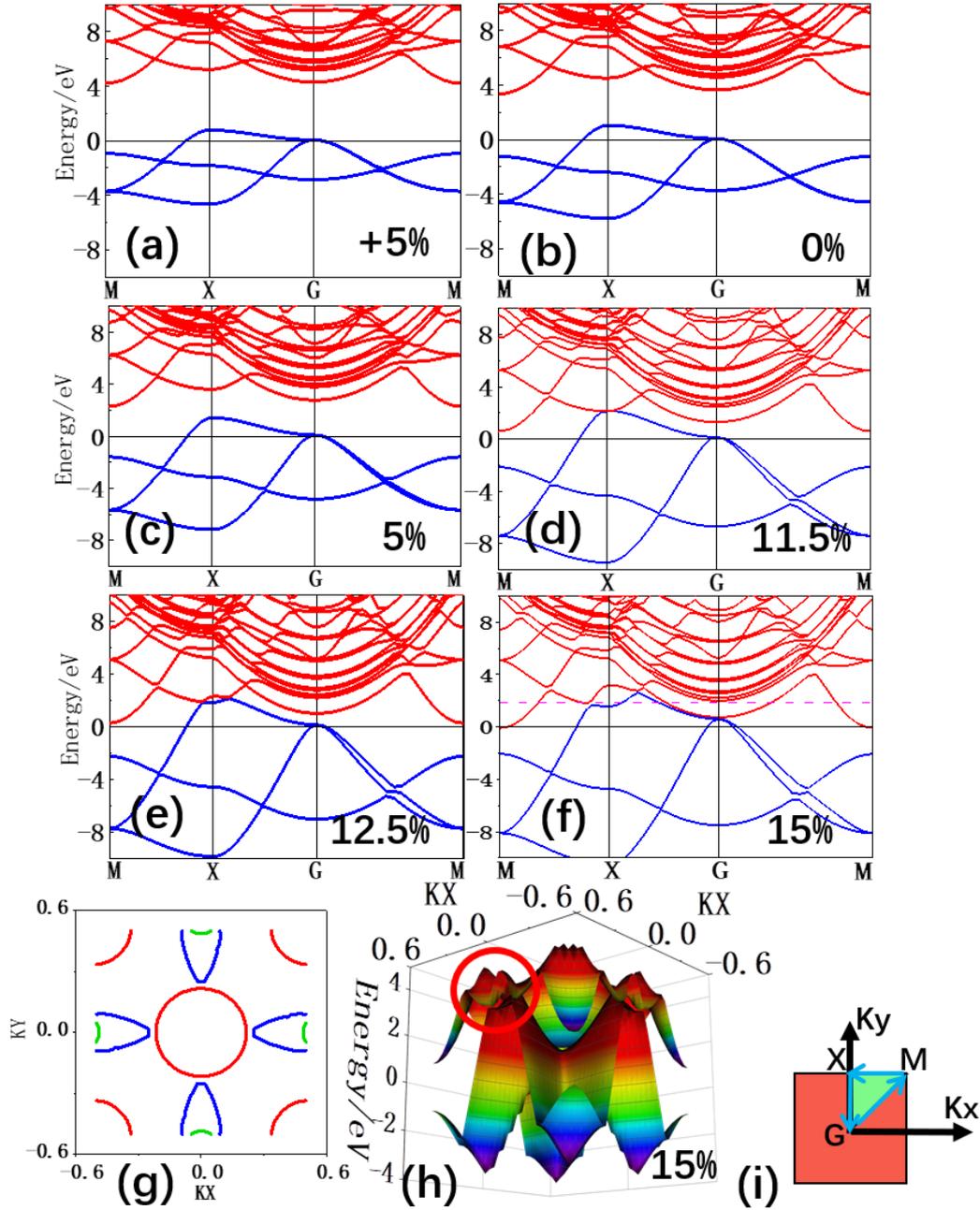

**Fig. 2** Band structure of iodiene under biaxial strain of (a): 5% tensile strain, (b): 0% without strain, (c): 5% compressive strain. (d): 11.5% compressive strain, (e): 12.5% compressive strain, and (f): 15% compressive strain. (g): Cross section of 3D band structure along 2.5 eV according to (f) and (h). The red is the conduction band section, and the blue is the valence. See the dotted line at (f) for section position. (h):3D band structure at 15% compressive strain. (i): The orientation of the integral path in the momentum space of the BZ.

The 3D band structures have demonstrated the band inversion structure in detail (Fig. 3). Since applying compressive strain in real space is just like exerting compressive force between CB and VB in BZ, increasing compressive strain can make HOMO and LUMO closing to zero. After that, further strain can make the band inversion both in VB and CB occur at the contact point in BZ. When the compression strain reaches 12.5%, the band inversion occurs both in HOMO and LUMO simultaneously at the four high symmetry points X in BZ (Fig. 3a$_1$, and inset), where the inset illustrates the red-circled inversion part in high symmetry point X. As for the valence band inversion, the 3D HOMO (Fig. 3a$_2$, and inset) shows the bands have inverted to a concaved structure, implying the inversion happens.

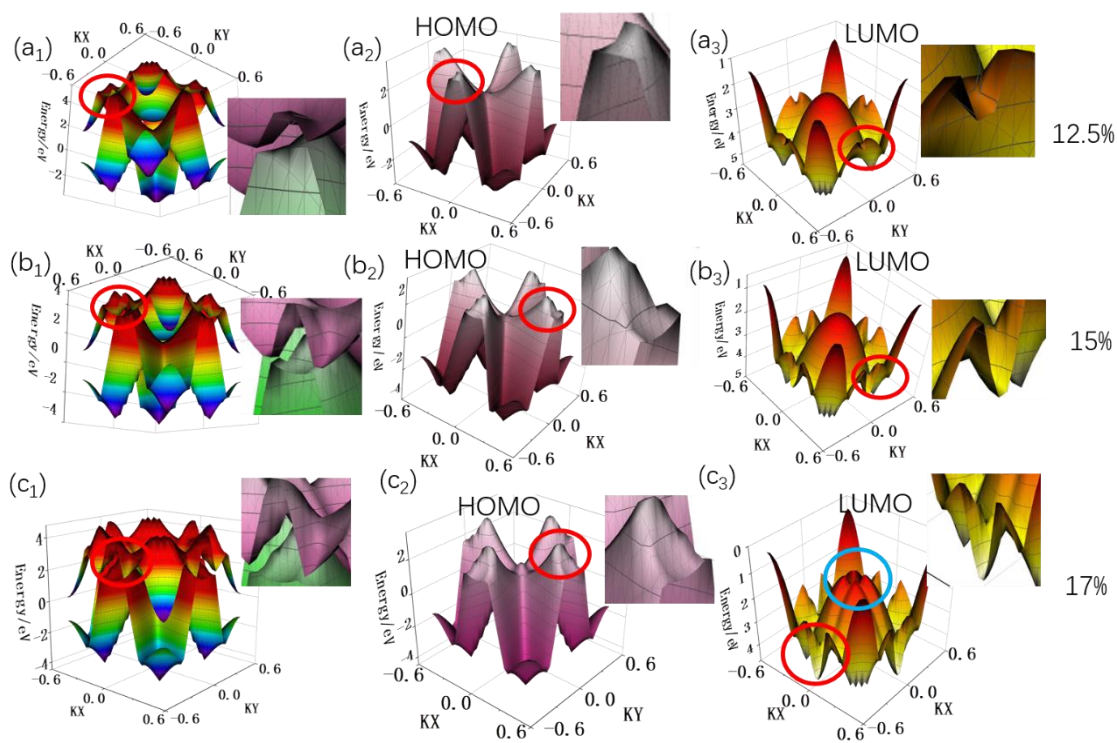

**Fig. 3** Iodiene 3D band structure and its constituent parts, including HOMO, LUMO, and related band inversion (red circle) under biaxial strain of (a$_1$-a$_3$): 12.5% biaxial stress, (b$_1$-b$_3$): 15% biaxial stress, and (c$_1$-c$_3$): 17% biaxial stress. The red circle points out the inversion part, enlarging as inset. The blue circle marks the inversion configuration at gamma point to LUMO (c$_3$), inverting 180° to its back in comparison to its normal structure.

The LUMO structure has been turned 180° to its back (Fig. 3a$_3$, and inset), the circled area and inset have marked the inversion structure. With the strain increasing to 15% (Fig. 3b$_1$, and inset), the band inversion becomes even stronger than that in strain of 12.5% both in HOMO (Fig. 3b$_2$, and inset) and LUMO (Fig. 3b$_3$, and inset). All the inversion happens at the high symmetry points X, instead of point Gamma. In addition, a new contact happens in Gamma point (Fig. 3b$_3$, and inset). Thus, further compressive strain can make a new band inversion occurs in Gamma point (Fig. 3c$_1$, and inset), as well as in X points (Fig. 3c$_2$, and inset). The LUMO structure has been demonstrated as its back (Fig. 3c$_3$, and inset), the inverted band at X point has illustrated as red circle and inset, and that at gamma point in blue circle (Fig. S3).

When iodiene is subjected to uniaxial strain, a saddle shape band configuration has been observed (Fig. 4, Fig. S4), suggesting a low symmetry is available. The four equivalence peaks have changed to

"strong" and two "weak" in comparison to the biaxial strain (Fig. 4a$_2$-c$_2$, Fig. S4). The band structure along weak peak path "G-M-X" (Fig. 4a$_1$-c$_1$, Fig. 3d) is much different from that along strong one along "G-M-E" (Fig. 4a$_3$-c$_3$, Fig. 4f).

When compression strain proceeds along *a*-axis (Fig. 4g), the band gap reduce with the stain increasing (Fig. 4a$_1$-a$_3$). A gap remains kept until the stain is 12.5%, and achieves zero at gamma point and E point simultaneously when the strain reaches 14% (Fig. 4b$_1$-b$_3$). The band inversion occurs at the gamma point and E point simultaneously when strain has increased to 17% (Fig. 4c$_1$-c$_3$), making the Dirac points or topological nontrivial points available. The topological nontrivial points can connect into line (Fig. 4e, Fig. S4), demonstrating as close hoop (red) near gamma point (Fig. 4e, Fig. S4). The effect of SOC on iodiene is to create a distance between its valence bands, but it has no effect on the band inversion, suggesting the topological nontrivial point depends on specific condition (Fig.S5) . The valence band section is in blue and the conductance band section is in red. the band structure has changed from C4 symmetry to C2 symmetry.

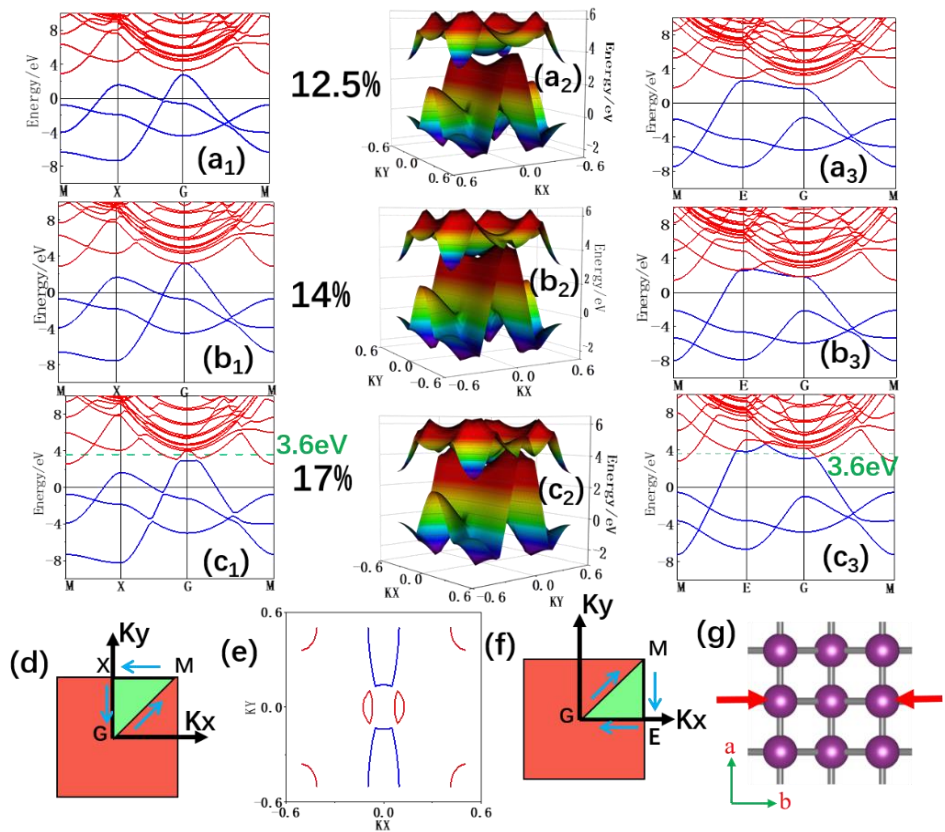

**Fig. 4** Band structure of iodiene under uniaxial strain of (a$_1$):12.5% strain, (b$_1$):14% strain, (c$_1$):17% strain along (d): the first path underlying high symmetry points in BZ. The 3D band structure by compressive strain of (a$_2$):12.5% strain, (b$_2$):14% strain, (c$_2$):17% strain, and (a$_3$-c$_3$)2D band structure along 2$^{nd}$ path way in BZ (e): Cross-section of 3D band structure at 3.6 eV of (c$_1$) and (c$_3$). (f): The second path along high symmetry points in BZ. (g): Schematic diagram of applying uniaxial strain.

## 3. Orbital hybridization

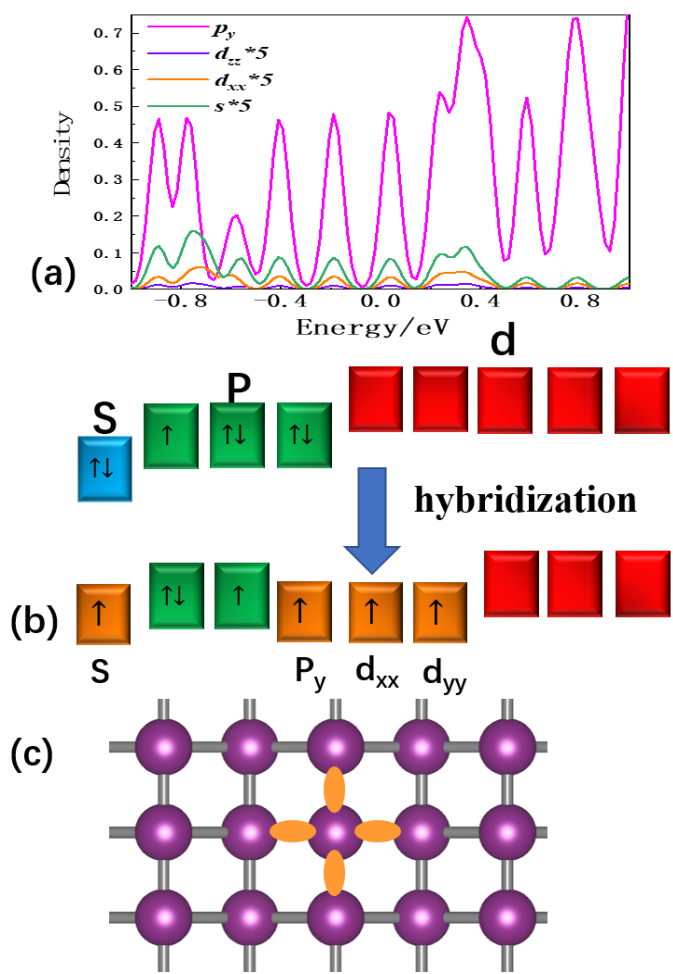

**Fig. 5** The hybridization orbit of iodiene. (a): Partial density of state (PDOS) of iodiene, suggesting the state density near the Fermi surface is contributed by four orbitals $spd^2$: $P_y$, $S$, $d_{ZZ}$ and $d_{XX}$. (b): Schematics of orbital levels and orbit hybridization. The orbital hybridization has contribution coming from $d$ orbit, accommodating the electrons from $S$, $P_y$ orbits. (c): Schematic diagram of the hybridization orbit in iodiene, and the yellow bonds represent the 4 σ bonds, arising from hybridization orbits.

The band hybridization can refer to Fig. 5, where the $spd^2$ hybridization orbitals have been formed, which is composed of σ bonds. Four covalent σ bonds can form the frame plane of iodiene (FPI), sustaining the stability. While the unpaired electron tends to form π bond, depending on the bond length. When the bond length is within van der Waals diameters (4.30Å)[33], a covalent will available. When the bond is around 3.0Å, the band inversion can be observed, inducing the π bond survival.

The $spd^2$ hybridization orbit is coming from the almost equal energy level (Fig. S6) over $s$, $p$, $d$ at the same period in iodine atom. The $s$ wave and $p$ wave have the most important contribution and $d$ have much weaker contribution (Fig. 5a), however, the covalent connection is performed to sustain the system.

The hybridization energy level diagram (Fig. 5b) illustrate the hybridization orbits (Fig. 5b yellow), the primitive *p* wave orbits (Fig. 5b green), and empty orbit (Fig. 5b red). The spatial configuration can refer to Fig. 5c, where the unpaired electron and single paired electrons' orbit mount in opposite side of the FPI plane. The unpaired electrons are active and dangling bond, implying a potential π band. If the bond length become much close, the weak antilocalization activates the electrical conductance, implying the band gap approaching zero. If the compress strain is strong enough to trigger the interaction between two contiguous lone electrons pair, the lone paired electrons' orbit can be activated, and making iodiene become buckled and behaving like that in BP. However, here only the planar configuration is available, suggesting the lone electrons pair have no effective interaction with its adjacent one.

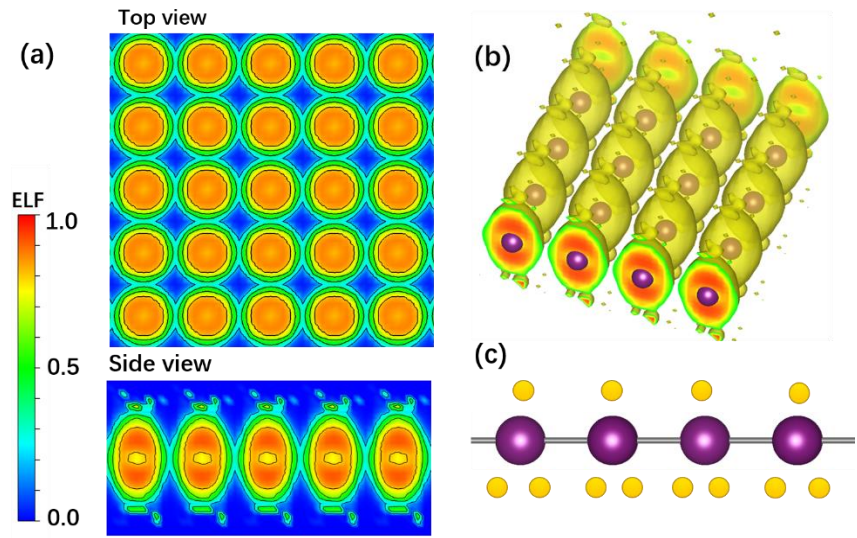

**Fig. 6** Electron local function (ELF) of iodiene. (a): Top-view and side view of ELF diagrams demonstrate that the bonding mode of iodiene has covalent properties. (b): Electron local function diagram, mimicking egg (ellipsoid) because of the covalence interaction between the contiguous atoms is much different from (c) the dangling unpaired electron and lone paired one.

Electronic local functions (ELF)[37] is presented to demonstrate the binding energy between the atoms (Fig. 6). From blue to ultra-red, the corresponding value means the covalent bond become stronger gradually (Fig. 6 a legend). Here the ELF value is around 0.2-0.6 (Fig. 6 a top view), implying a weak bond level in comparison to graphene[38]. Moreover, a side view (Fig. 6a) and the 3D images (Fig. 6b) have confirm the atoms in iodiene is connected by covalent bond with bond length as 3.234Å, much less than Van der Waals diameter, suggesting the covalence bond is strong enough to support the 2D configuration (Fig. 6a bottom side view). The ellipse shape, instead of circle, represents the deformed shape is arising from strong covalence bond, and the difference between covalence bond and dangling bonds (Fig. 6b and Fig. 6a bottom side view). The interaction between the atoms include three parts. (1) σ-σ bonds (2) π–π bonds and (3) lone pair-lone pair interaction. Here, only σ-σ bond works. The lone unpaired electron represents π–π bond if LBL is suitable. With the compressive strain applying, the π–π bond become active finally. If a severe strain is applied, the lone pair-lone pair interaction might run.
.

## Summary

The 2D materials, iodiene, is really existing in extreme high pressure. Moreover, after DFT optimization, the iodiene can be available, even if in atmosphere pressure. Our findings have opened a new way to 2D crystal by applying hybridization rule, where the same period *d* wave orbit can hybridize with *s* wave and *p* wave.

## Methods

First-principles calculations are performed by the plane wave code Vienna ab initio simulation package (VASP)[39], The calculations were carried out within the Generalized-gradient approximation (GGA) with the Perdew−Burke−Ernzerh of functional (PBE) to the density-functional theory (DFT)[40]. The wave functions between the cores are expanded in plane waves with a kinetic energy cutoff of 400 eV. The convergence criteria of the electron self-consistent loop were set to $10^{-8}$ eV and 0.01 eV/Å for the force convergence criterion. The K-mesh scheme is Gamma and use 0.03 2π/ Å accuracy for K point sampling in the reciprocal space. To prevent the interaction of periodic structures, a 30-angstrom vacuum layer was used. The phonon spectrum is calculated by DFPT[41] algorithm with 2×2×1 supercell. According to the formula

$$E_b = E_{at} - E_{sheet} \qquad (1)$$

the binding energy of materials ($E_b$) are calculated[42], where $E_{at}$ is the energy of an isolated spin polarized atom and $E_{sheet}$ is the energy of each atom in the two-dimensional materials.

By multiply the lattice constants a and b by the same scaling coefficient, the biaxial strain is implemented. For the primitive cells of p4mmm space group, the lattice constants a and b are equivalent and the calculation of the uniaxial strain is to multiply one of the lattice constants by the corresponding scaling coefficient.

## Acknowledgment

This work was supported partly by grants from the National Natural Science Foundation of China (No. 51771125 and 51472170). We also thank Dr. Ren for his help with new version of VASP software.

## Table of Contents Entry

The 3D iodine molecule crystalline is stacked by 2D molecule layers, transferring into covalent-connection atomic layer in-plane and Van der Waals connection inter-plane by pressure-inducing process, suggesting 2D iodiene can be really survival, and the lone unpaired electron is possible to form π key if bond length is reduced effectively.

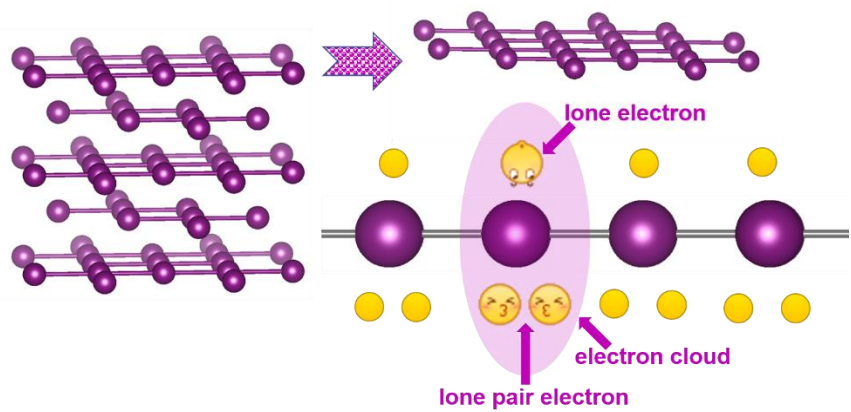

## Keywords

Iodine, Two-dimension materials, Planar, Density functional calculations, First-principles calculations, Strain, Band structure, Electrical properties

## Author Contributions

Q. S. H conceived the project, and perfromed the main analysis of Data . X. Y. Z performed the caculation and simulation. All authors discussed the results, and wrote the paper.

# Supporting information

# Towards planar Iodine 2D crystal materials

Xinyue Zhang, Qingsong Huang*

School of chemical engineering, Sichuan University, Chengdu, 610065, P. R. China

**1. Puckered configuration**

Puckered configuration of iodiene (Fig. S1 (a)) is unstable, provided the geometric optimization should transfer the puckered iodiene into planar configuration. In addition, the phonon spectra demonstrate a negative branch, implying a configuration might not be sustained stably.

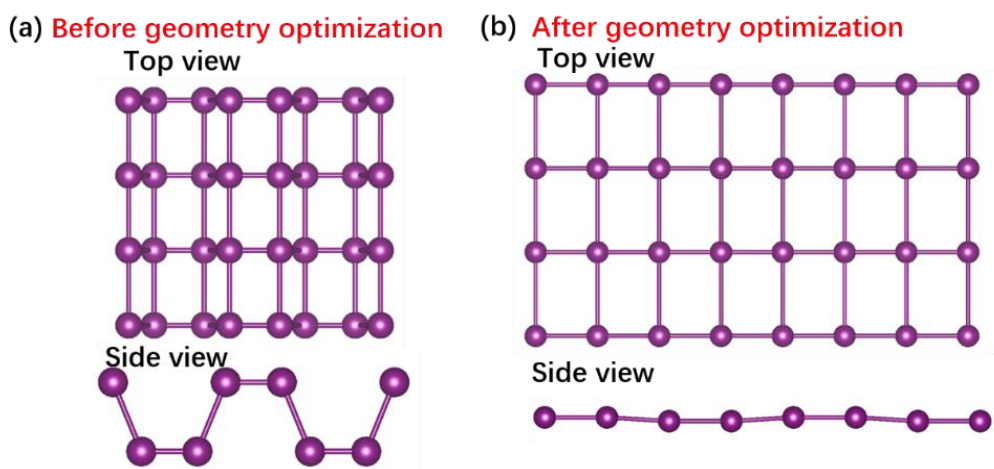

**Fig. S1** Puckered configuration of iodiene before and after geometric optimization. (a): puckered iodiene before geometric optimization. (b): puckered iodiene after geometric optimization. The geometrically optimized iodiene shows a quasi-plane structure, indicating that the puckered structure of iodiene cannot exist stably.

* School of chemical engineering, Sichuan University, Chengdu, 610065, P. R. China
  E-mail: qshuang@scu.edu.cn
  (Prof. Qingsong Huang)

## 2. Rectangle lattice (a≠b)

According to the compression ratio of iodine crystal under high pressure[1] (a:0.9406, b:0.9329), the compression strain of iodiene under normal pressure was performed. There is no virtual frequency in iodiene phonon spectrum, which indicates that iodiene is robust of in-plane compression strain.

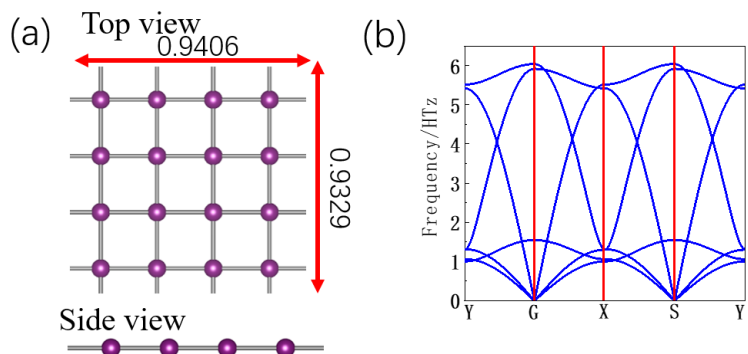

**Fig.S2** Iodiene is stable under strain. (a): The a and b axes of iodiene are adjusted according to the strain ratio of iodide crystal under high pressure. (b): There is also no virtual frequency in the iodiene phonon spectrum with varying ratio of a and b, which proves that the iodiene has certain robustness to strain.

## 3. Correlation of bi-axial compression strain in real space to pushing closing force between HOMO and LUMO in BZ

The 3D band structure of iodiene under strain was studied. and it was found that under biaxial strain, the valence band gradually increased, and when the strain reached 11.5%, the valence band and the conduction band were tangent at X point. Further compressive strain behaves like exerting pushing-close force to make the LUMO and HOMO band closer, becoming inversion structure after HOMO contacting with LUMO.

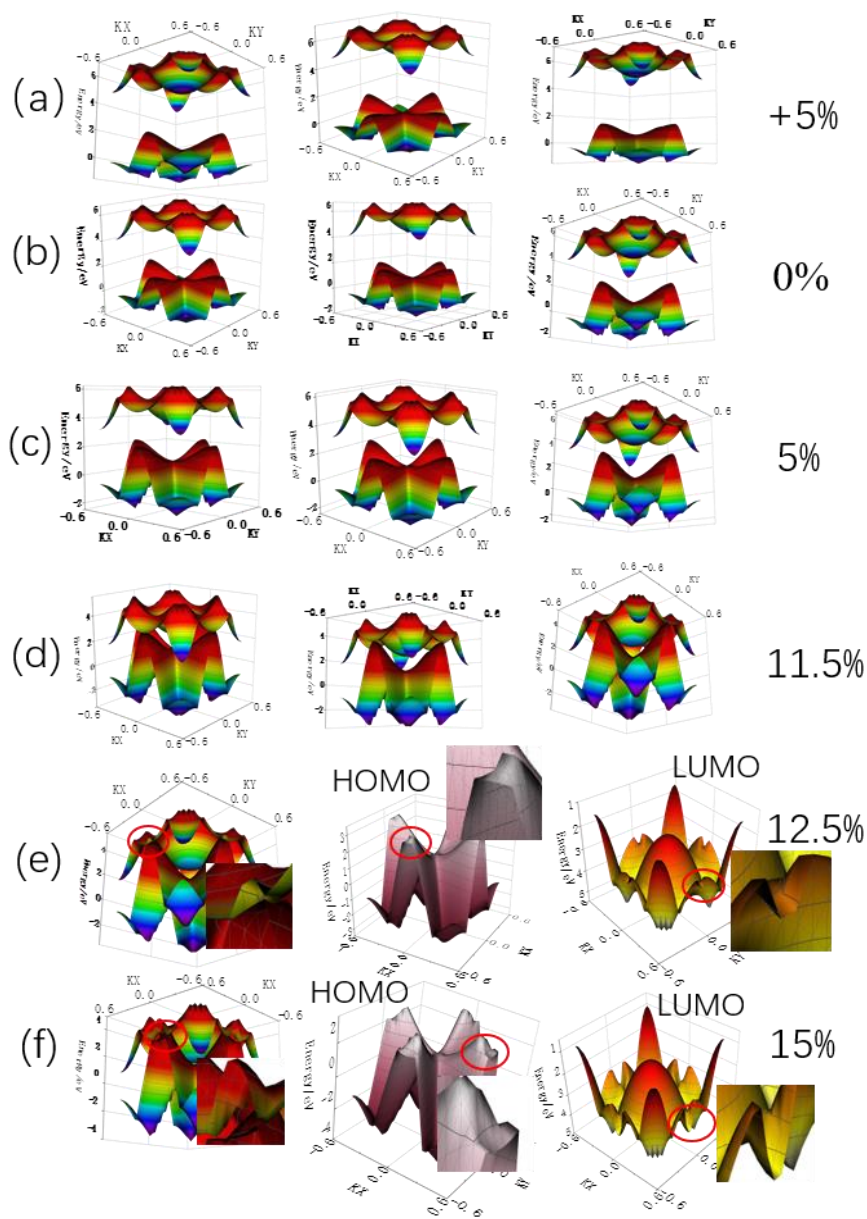

**Fig. S3** 3D band structure of iodiene from under different strain. (a-d): Band structure are viewed from three different perspectives. (a):5% tensile strain. (b): Without strain. (c):5% compressive strain. (d):11.5% compressive strain. (e, f): The 3D band structure and the HOMO, LUMO 3D band structure. (e):12.5% compressive strain. (f): 15% compressive strain.

## 4. Suffering Uni-axial strain

The uni-axis strain has the same effective on the band structure as bi-axial strain, except the symmetry of 3D band structure should be broken by the asymmetry strain on the lattice. The effect of uniaxial strain on iodiene band structure is similar to that of biaxial strain. With the increase of compression strain, the valence band rises continuously (Fig.S4 a). HOMO and LUMO have just flipped at X point when the uniaxial compression strain reaches 14% (Fig. S4 b). Further increasing the strain to 17% (Fig.S4 c), the contact parts of the HOMO and LUMO reversed. However, due to the decrease in symmetry caused by uniaxial strain, the morphology of band inversion is different from that of biaxial strain.

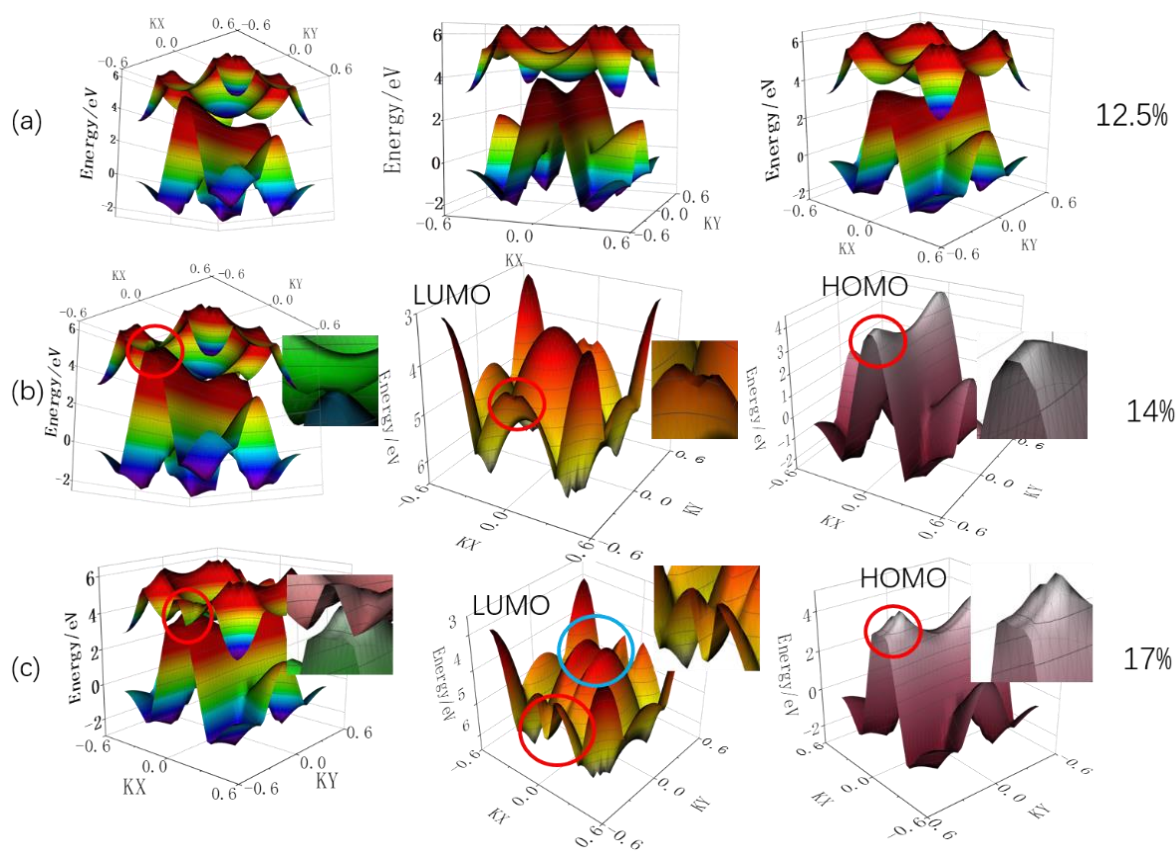

**Fig. S4** Modulation of 3D band structure by uniaxial strain. (a):Band structure under 12.5% uniaxial compression strain are viewed from three different perspectives. (b),(c):The 3D band structure and the HOMO,LUMO 3D band structure. (b):14% uniaxial compression strain. (c):17% uniaxial compression strain.

## 5. Band structure correlating between with SOC and without

The effect of SOC on band structure of iodiene is to create a distance between valence bands of iodiene, but SOC has no effect on the band inversion.

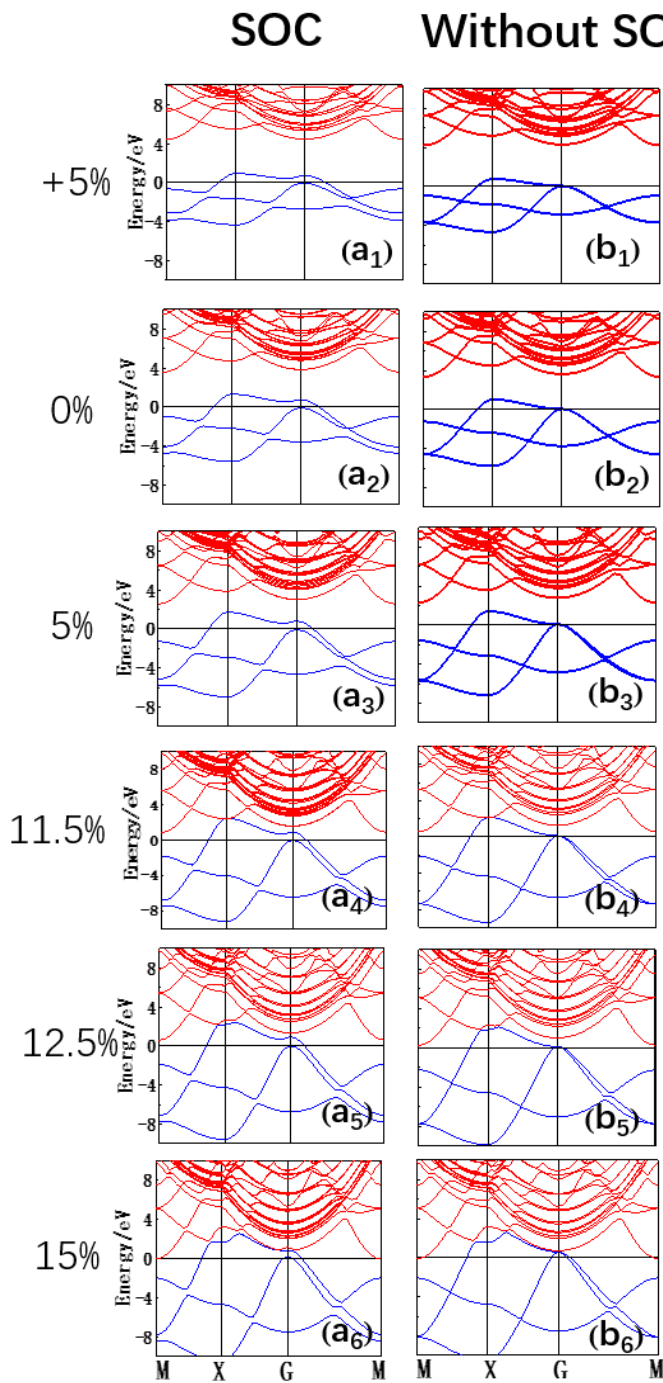

**Fig. S5** Effects of SOC and strain on iodiene band structure. ($a_1$-$a_6$):Band structure of iodiene under SOC effect. ($b_1$-$b_6$):Band structure of iodiene without SOC effect. ($a_1$, $b_1$):With 5% tensile strain. ($a_2$,$b_2$):Without strain. ($a_3$,$b_3$):With 5% compression strain. ($a_4$,$b_4$):With 11.5% compression strain. ($a_5$,$b_5$): With 12.5% compression strain. ($a_6$,$b_6$):With 15% compression strain.

## 6. PDOS of elements

As is shown in Fig.S5, the DOS of iodine is stretched over four energy levels, occupying all the orbits including the *s*, *p*, and *d*. Each orbital has a different contribution on the four energy levels, offering some specific overlap of energy level with each other. This overlap provides the conditions for orbital hybridization.

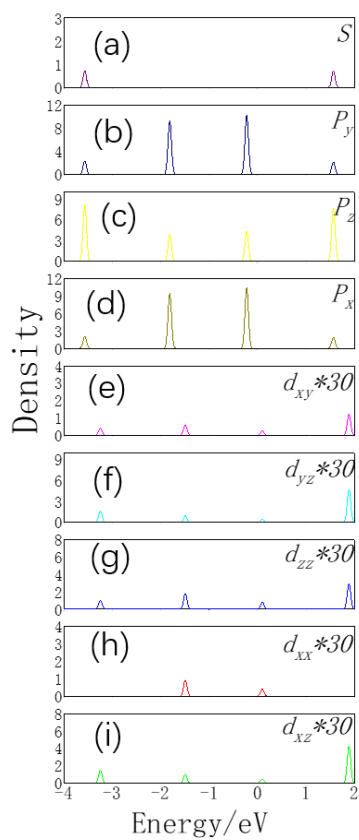

**Fig. S6** *Pdos* of iodine atoms. (a):*s* orbit. (b):$p_y$ orbit. (c):$p_z$ orbit. (d):$p_x$ orbit. (e):30*$d_{xy}$ orbit. (f):30*$d_{yz}$ orbit. (g):30*$d_{zz}$ orbit. (h):30×$d_{xx}$ orbit. (i):30×$d_{xz}$ orbit.

**Basic parameter list**
**Table. S1** The properties of iodiene: total energy ($E_{total}$), binding energy ($E_b$), lattice constant (l), bond length (b), bond angle (θ), bandgap ($E_{gap}$), and space group

| $E_{total}$ | $E_b$* | l | b | θ | $E_{gap}$ | space group |
|---|---|---|---|---|---|---|
| -1.28 eV | 1.238 eV | 3.234 Å | 3.234 Å | 90° | 3.470 eV | P4/MMM |

* Binding energy is calculated according to formula (1).

The total energy is negative, and the binding energy is positive, proving the stability of iodiene. Since there is only one atom in the original cell, so the bond length is equal to the lattice constant, 3.234 Å, with broadband gap of 3.470 eV. Iodiene belongs to P4/MMM space group with bond Angle of 90°.